# Convergence of Optimal Expected Utility for a Sequence of Discrete-Time Markets

David M. Kreps and Walter Schachermayer[*]

February 5, 2020

**Abstract**

We examine Kreps' conjecture [17] that optimal expected utility in the classic Black–Scholes–Merton (BSM) economy is the limit of optimal expected utility for a sequence of discrete-time economies that "approach" the BSM economy in a natural sense: The $n$th discrete-time economy is generated by a scaled $n$-step random walk, based on an unscaled random variable $\zeta$ with mean zero, variance one, and bounded support. We confirm Kreps' conjecture if the consumer's utility function $U$ has asymptotic elasticity strictly less than one, and we provide a counterexample to the conjecture for a utility function $U$ with asymptotic elasticity equal to 1, for $\zeta$ such that $E[\zeta^3] > 0$.

## 1 Introduction

Fix a random variable $\zeta$ with mean zero, variance one, and bounded support. For $n = 1, 2, \ldots$, construct a financial-market economy with two securities, a riskless *bond*, which serves as numeraire (hence, has interest rate 0) and a risky security, called the *stock*, which trades against the bond in frictionless markets at time $0, 1/n, 2/n, \ldots, (n-1)/n$. The price process for the stock is generated as follows: For an i.i.d. sequence $\{\zeta_j; j = 1, 2, \ldots\}$, where each $\zeta_k$ has the distribution of $\zeta$, the law for the price of the stock at time $k/n$ is

$$S(k/n) := e^{\xi(k/n)} \quad \text{where} \quad \xi(k/n) := \sum_{j=1}^{k} \frac{\zeta_j}{\sqrt{n}}.$$

($\xi(0) \equiv 0$ and $S(0) \equiv 1$.) At time 1, the bond pays a consumption dividend of 1, and the stock pays a consumption dividend of $S(1)$ defined as above.

Embed this model into the standard state space $\Omega = C_0[0, 1]$. Let $\omega$ denote a generic element of $\Omega$, with $\omega(t)$ the value of $\omega$ at time $t$. Endow $\Omega$ with the sup-norm topology;

[*](Stanford Graduate School of Business and University of Vienna, Faculty of Mathematics, respectively.) Kreps' research is supported by the Graduate School of Business, Stanford University. Schachermayer gratefully acknowledges support by the Austrian ScienceFund (FWF) under grant P28661 as well as by the Vienna Science and Technology Fund (WWTF) through projects MA14-008 and MA16-021. Schachermayer also thanks the Department of Mathematics, Stanford University, for its hospitality while this paper was being written.

let $F$ denote the Borel $\sigma$-field, and let $\{F_t; t \in [0,1]\}$ denote the standard filtration on $\Omega$. For each $n$, let $\mathcal{P}_n$ be the probability measure on $\Omega$ such that the joint distribution of $(\omega(0), \omega(1/n), \ldots, \omega(1))$ matches the distribution of $(\xi(0), \xi(1/n), \ldots, \xi(1))$, and such that $\omega(t)$ for $k/n < t < (k+1)/n$ is the linear interpolate of $\omega(k/n)$ and $\omega((k+1)/n)$. And let $S : \Omega \to R_+$ be defined by $S(\omega, t) = e^{\omega(t)}$.

Donsker's Theorem tells us that $\mathcal{P}_n \Rightarrow \mathcal{P}$, where $\mathcal{P}$ is Wiener measure on $C_0[0,1]$; that is, $\omega$ under $\mathcal{P}$ is a standard Brownian motion, starting at $\omega(0) = 0$, and $S(\omega)$ under $\mathcal{P}$ is geometric Brownian motion, so that $\mathcal{P}$, together with the riskless bond, prescribes the simple continuous-time economy of Black and Scholes [5] and Merton [21] (hereafter, the BSM economy or model).

We imagine an expected-utility-maximizing consumer who is endowed with initial wealth $x$, with which she purchases an initial portfolio of stock and bond. Thereafter, she trades in non-anticipatory and self-financing fashion in the stock and bond (that is, (a) the information she possesses at time $k/n$, on which basis she trades, is [only] the history of the stock price up to and including time $k/n$, and (b) any purchase of stock after time 0 is financed by the sale of bonds, and the proceeds of any sale of stock are invested in bonds), seeking to maximize the expectation of a utility function $U : (0, \infty) \to R$ applied to the final dividend generated by the portfolio she holds at time 1.

The question that forms the basis for this paper is: If we place this consumer in the $n$th discrete-time economy (where the stock and bond trade (only) at times $0, 1/n, 2/n, \ldots, (n-1)/n$), does the optimal expected utility she can attain approach, as $n \to \infty$, what she can optimally attain in the continuous-time BSM economy?

Let $u_n(x)$ be the supremal expected utility she can attain in the $n$th discrete-time economy if her initial wealth is $x$, and let $u(x)$ be her supremal expected utility in the BSM economy. Kreps [17] obtains partial one-sided results, showing that $\liminf_n u_n(x) \geq u(x)$. And he proves $\lim_n u_n(x) = u(x)$ in the very special cases of $U$ having either constant absolute or relative risk aversion. But he only conjectures that the second "half", or $\limsup_n u_n(x) \leq u(x)$ is true for gernal (concave and differentiable) $U$.

Employing the notion of *asymptotic elasticity of utility* from Kramkov and Schachermayer [16] (and making extensive use of their analysis), we verify that $\lim_n u_n(x) = u(x)$ if the utility function $U$ has asympototic elasticity less than 1. However, we show by example that if the asymptotic elasticity of $U$ is 1, it is possible that $u(x)$ is finite while $\lim_n u_n(x) = \infty$, both for all $x > 0$.

## 2 Previous and Contemporaneous Literature

A substantial body of literature concerns utility maximization problems in financial markets, going back to seminal work by R. Merton [20] and continuing, e.g., in ([6, 13, 16, 8, 11, 24, 4]). As regards the continuity of utility maximization under weak convergence of financial markets, positive results are obtained in ([10], [23], [25]); these results all assume that, in each discrete-time model, markets are complete.

Our interest, motivated by the discussions in [17], is in cases where the discrete-time markets are incomplete. Since the seminal paper of Cox, Ross, and Rubinstein [7], financial economists have believed that, if $\zeta$ has two-element support (the so-called binomial case), and so markets are complete in each discrete-time economy, then these discrete-time economy for large $n$ behave (in economic terms) like the continuous-time limit, at least for the BSM continuous-time limit. But what if $\zeta$ has support of, say, size three, but there are only the two securities? Markets are incomplete for any finite $n$; does this incompleteness mean very different economic outcomes? Or, if the probability laws $\mathcal{P}^n$ that govern the discrete-time security-price processes converge weakly to $\mathcal{P}$, is it then true that $\lim_n u_n(x) = u(x)$?

It is already known that weak convergence is insufficient. Merton [20] observes that if $U$ has constant relative risk aversion with risk-aversion parameter less than $1/2$, the optimal strategy in the BSM economy is to short-sell bonds, leveraging to achieve a (fixed) fraction greater than 100% of current wealth in the risky asset. Suppose that, in our special discrete-time setting, where the security-price process is driven by scaled copies of a single random variable $\zeta$, $\zeta$ has support that is unbounded below. Trying to achieve such a leverage strategy in any of the finite-time economies would give a positive probability of bankruptcy, which is incompatible with these utility functions. The best an investor can do for large enough $n$ in these circumstances is to hold 100% of her wealth in the risky asset, which results in $\lim_n u_n(x) < u(x)$.

On the other hand, weak convergence of $\mathcal{P}^n$ to $\mathcal{P}$ alone does not preclude the possibility of asymptotic arbitrage ([12], [15]), in which case $\lim_n u_n(x) = \infty$, even when $u(x)$ is finite valued (and $U$ is very well behaved); see [17], Chapter 7.

By assuming in our setting that $\zeta$ has bounded support, we avoid the first problem. And, in our setting, asymptotic arbitrage is precluded; see [17], Proposition 7.1. Still, ill-behaved $U$ can pose problems: Within our setting, we show that $\lim_n u_n(x) = u(x)$ for all $x > 0$ if $U$ has asymptotic elasticity less than 1. But if $U$ has asymptotic elasticity of 1, even if markets are complete for each $n$, convergence can fail, and fail in spectacular fashion.

The incomplete-market case on which we focus has recently been treated in a setting of greater generality by E. Bayraktar, Y. Dolinsky, and J. Guo [2].[1] Their paper assumes that the financial markets $(S^n)_{n=1}^\infty$ are general semi-martingales and the limiting market $S$ is a continuous semi-martingale. Also, the utility function $U$ in [2] may measurably depend on the observed trajectory of the stock price. Hence their model includes our special and paradigmatic case, where $(S^n)$ is induced by a single (scaled) random variable $\zeta$ and $S$ is geometric Brownian motion. In this more general setting, they make assumptions sufficient to show that $\lim_n u_n(x) = u(x)$. The key assumptions in [2] are Assumption 2.3 (ii), that a certain family of random variables is uniformly integrable, and Assumption 2.5, which effectively assumes away the possibility of asymptotic arbitrage. Lemma 2.2 of [2] provides some fairly strong conditions under which Assumption 2.3 (ii) is satisfied, conditions that are not related to the concept of asymptotic elasticity. In comparison, we *deduce* the uniform integrability of certain corresponding families of dual random variables (see (8.8) and (8.14) below) from the assumption that $U$ has asymptotic elasticity less than 1. And, in our more limited setting, the impossibility of asymptotic arbitrage is a conclusion.

---

[1] This paper [2] was put on ArXiv in November 2018; the authors kindly brought their paper to our attention after a first version of the current paper appeared on ArXiv in July 2019. The references to their results refer to the ArXiv version of [2] from September 2019.

# 3 The utility function, its conjugate function, and asymptotic elasticity

We always assume the following:

***Assumption (3.1).*** *The utility function $U$ is strictly increasing, strictly concave, and continuously differentiable, and satisfies the Inada conditions that $\lim_{x\to 0} U'(x) = \infty$ and $\lim_{x\to\infty} U'(x) = 0$. Moreover, without loss of generality and for notational convenience later, we assume unless otherwise specified that $U$ is normalized so that $\lim_{x\to\infty} U(x) > 0$, without precluding the possibility that $\lim_{x\to\infty} U(x) = \infty$. (Of course, $\lim_{x\to 0} U(x)$ can be either finite or $-\infty$.)*

We let $V$ denote the conjugate function to $U$: For $y > 0$,

$$V(y) = \sup_{x>0} \big[U(x) - xy\big], \quad \text{for } y > 0.$$

The following results are standard (see, e.g., [16]) and follow from Assumption (3.1):

- Let $I : (0,\infty) \to (0,\infty)$ be the inverse of $U'$; that is, $I(y) = (U')^{-1}(y)$. Then for every $y \in (0,\infty)$, $V(y) = U(I(y)) - yI(y)$.

- The function $y \to V(y)$ is strictly convex, continuously differentiable, and strictly decreasing.

- $V(\infty) = U(0)$ and $V(0) = U(\infty)$, where the values of $U$ and $V$ at 0 and $\infty$ are interpreted as the limits as $x$ and $y$ approach 0 and $\infty$, respectively.

- $V'(y) = -I(y)$, so $\lim_{y\to 0} V'(y) = -\infty$ and $\lim_{y\to\infty} V'(y) = 0$

- $U(x) = \inf_{y>0} \big[V(y) + xy\big], \quad \text{for } x > 0.$

The notion of *asymptotic elasticity* of $U$, defined in [16], plays an important role in our analysis. For the utility function $U$, its asymptotic elasticity, written $\text{AE}(U)$, is defined by

$$\text{AE}(U) := \limsup_{x\to\infty} \frac{xU'(x)}{U(x)}$$

If, for instance, $U(x) = x^\alpha/\alpha$ for $\alpha \in (0,1)$, then $\text{AE}(U) = \alpha$.

The concavity of $U$ implies that $\text{AE}(U) \leq 1$ in all cases; if $U$ is bounded above and if $U(\infty) > 0$, then $\text{AE}(U) = 0$. But if $U(\infty) = \infty$, $\text{AE}(U)$ can equal 1; an example is where $U(x) = x/\ln(x)$ for sufficiently large $x$.[2]

Many of our results depend on the assumption that $\text{AE}(U) < 1$, which derives from a comparison of the average and marginal utilities provided by $U$ as the argument of $U$ approaches $\infty$: $\text{AE}(U) < 1$ is equivalent to:

$$\text{For some } \gamma < 1, \quad U'(x) < \gamma \frac{U(x)}{x}, \quad \text{for all large enough } x.$$

---
[2]The conditions (3.1) on $U$ include $\lim_{x\to\infty} U(x) > 0$; this is solely so that $\text{AE}(U) \geq 0$ in all cases.

The economic interpretation of this condition can be sharpened if we think of the consumer comparing her marginal utility from very large consumption levels with the average utility she has accrued over some base level $x_0$. Because

$$\lim_{x \to \infty} \frac{U(x) - U(x_0)}{x - x_0} = \lim_{x \to \infty} \frac{U(x)}{x}, \text{ for all } x_0 > 0,$$

$\text{AE}(U) < 1$ is equivalently:

For some $\gamma < 1$ and every $x_0 > 0$, $U'(x) < \gamma \frac{U(x) - U(x_0)}{x - x_0}$, for all large enough $x$,

where "large enough" depends on the value of $x_0$.

As noted in [26], the concept of asymptotic elasticity connects to the limiting behavior of relative risk aversion by an application of de l'Hôpital's rule as follows: If the limit of the coefficient of relative risk aversion, $\lim_{x \to \infty} -xU''(x)/U'(x)$, exists and is strictly positive, then $\lim_{x \to \infty} xU'(x)/U(x)$ exists and is less than 1; that is, $U$ has asymptotic elasticity less than 1. Since it is believed to be "common" for economic agents to have non-increasing relative risk aversion, this belief implies that agents with this common property have asymptotic elasticity less than one.

## 4 Solutions for the continuous-time economy

As is well known, the continuous-time BSM economy admits a unique equivalent martingale measure denoted by $\mathcal{P}^*$; that is, a probability measure on $\Omega$ that is probabilistically equivalent to $\mathcal{P}$ and such that $\{S(\omega, t); t \in [0, 1]\}$ is a martingale (over the natural filtration $\{F_t\}$). This measure $\mathcal{P}^*$ has Radon-Nikodym derivative with respect to $\mathcal{P}$ given by

$$\left(\frac{d\mathcal{P}^*}{d\mathcal{P}}\right)(\omega) = \exp\left(-\frac{\omega(1)}{2} - \frac{1}{8}\right).$$

And, as is well known, this economy has "complete markets." That is, the consumer can construct (as a stochastic integral) any measurable positive contingent claim $X$ that she can afford, where what she can afford is given by the single budget constraint $\mathbb{E}_{\mathcal{P}^*}[X] \leq x$, where $\mathbb{E}_{\mathcal{P}^*}[\cdot]$ denotes expectation with respect to $\mathcal{P}^*$. Hence, with wealth $x$, the consumer's problem is to

Maximize $\mathbb{E}_{\mathcal{P}}[U(X)]$, subject to $\mathbb{E}_{\mathcal{P}^*}[X] \leq x$.

Let

$$u(x) := \sup\left\{\mathbb{E}_{\mathcal{P}}[U(X)] : \mathbb{E}_{\mathcal{P}^*}[X] \leq x\right\}.$$

That is, $u(x)$ is the supremum of expected-utility level that the consumer can achieve in the BSM economy, starting with wealth $x$.

It is convenient for later purposes to define the density function $Z : C_0[0, 1] \to (0, \infty)$ by

$$Z(\omega) := \exp\left(-\frac{\omega(1)}{2} - \frac{1}{8}\right).$$

That is, $Z$ is the unique continuous (in $\omega$) version of the random variable $d\mathcal{P}^*/d\mathcal{P}$. Of course, $\mathbb{E}_{\mathcal{P}*}[X] = \mathbb{E}_{\mathcal{P}}[X \cdot Z]$ for any random variable $X$ such that (at least) one of the expectations makes sense. And, in this notation $u(x) = \sup \{\mathbb{E}_{\mathcal{P}}[U(X)] : \mathbb{E}_{\mathcal{P}}[X \cdot Z] \leq x\}$.

We have the following from Cox and Huang [6], Karatzas, Lehoczy, and Shreve [14], and [16]. (See, in particular, [16], Theorem 2.0.)

4.1. $u(x) \geq U(x)$ (since the consumer can always buy and hold $x$ bonds).

4.2. $x \to u(x)$ is continuously differentiable, strictly increasing, and concave.

4.3. From 4.1 and 4.2, if $u(x_0) < \infty$ for any $x_0 > 0$, then $u(x) < \infty$ for all $x > 0$.

4.4. If $u(x) < \infty$, *and if the consumer's problem has a solution (that is, if the supremum is attained)*, then there exists $y(x) > 0$ such that the solution has the form $X(\omega) = I(y(x) \cdot Z(\omega))$, where $y(x) = u'(x)$ and (as noted earlier) $I = (U')^{-1}$.

If the consumer's problem has a solution at wealth level $x > 0$, then it has a solution for all wealth levels $x' > 0$ such that $x' < x$.

However, it is possible that, at least for some $x$, $u(x) < \infty$ and yet the supremum that defines $u(x)$ is not attained by any contingent claim $X$. (An example is given in [16], Section 5; we produce examples below.) If (for the given utility function $U$) this is true for some finite $x$, let $\bar{x}$ be the infimum of all $x$ for which there is no solution (but $u(x) < \infty$); there is a solution at $\bar{x}$, and so the range of $x$ for which there is a solution is the interval $(0, \bar{x}]$.

The function $x \to u(x)$ is continuously differentiable and the "Lagrange multiplier function" $x \to y(x) = u'(x)$ is continuous and strictly decreasing on $(0, \bar{x})$.

4.5. Let $v$ be the conjugate function to $u$. That is,
$$v(y) = \sup\{u(x) - xy : x > 0\}, \quad \text{for } y > 0.$$
Then
$$v(y) = \mathbb{E}_{\mathcal{P}}[V(yZ)].$$
The function $y \to v(y)$ is convex and nonincreasing, and it is strictly decreasing and continuously differentiable *where it is finite*.

Of course, it may be that $u(x) \equiv \infty$, in which case $v(y) \equiv \infty$. But suppose $u(x) < \infty$ for some, and therefore for all, $x > 0$. While $u(x)$ is necessarily concave, differentiable, and strictly increasing, it is not in general true that $\lim_{x \to \infty} u'(x) = 0$. That is, the marginal (maximal expected) utility of wealth need not approach zero as the wealth level goes to $\infty$. Roughly speaking, this can happen when a consumer can purchase ever larger amounts of consumption on events of ever smaller probability, but where the ratio of the amount purchased to the probability of the event approaches infinity at a rapid enough rate. This idea was exploited by Kramkov and Schachermayer [16] for any utility function $U$ that satisfies $AE(U) = 1$, by choosing, based on $U$, specific measures that are different from but play an analogous role to $\mathcal{P}^*$ and $\mathcal{P}$. Here, $\mathcal{P}^*$ and $\mathcal{P}$ are fixed – they come from BSM – so we show this sort of possibility through the selection of specific utility functions $U$.

Admitting this is possible (we show that it is), consider the implications for $v$ and, in particular, for $v$ around the value $y_0$, where $y_0 = \lim_{x \to \infty} u'(x)$. To the left of $y_0$ ($y < y_0$), we have $v(y) = \infty$. To the right, $v(y)$ is finite. But what is the limiting behavior of $v$ as $y$ approaches $y_0$ from the right? In theory, we could have $\lim_{y \searrow y_0} v(y) = \infty$. Or $\lim_{y \searrow y_0} v(y) < \infty$, and $\lim_{y \searrow y_0} v'(y) = \infty$. Or $\lim_{y \searrow y_0} v(y) < \infty$, and $\lim_{y \searrow y_0} v'(y) < \infty$. All of these are possible. In fact, giving the full catalog, we have Proposition 1:

**Proposition 1.** *Assume that $U$ satisfies the conditions (3.1). It is possible that $u(x) = \infty$ (for all $x \geq 0$). But if $u(x) < \infty$ for some $x$, hence for all $x$, it must be that $x \to u(x)$ is strictly increasing. Moreover, we have the following possibilities.*

a. *For some utility functions $U$, $\lim_{x \to \infty} u'(x) = 0$, in which case $v(y)$ is finite for all $y > 0$. (Since $\lim_{y \to 0} v(y) = \lim_{x \to \infty} u(x)$ and $\lim_{y \to \infty} v(y) = \lim_{x \to 0} u(x)$, the function $v$ can have limit $\infty$ or a finite limit as $y$ approaches $0$; and $v$ can have limit $-\infty$ or a finite limit as $y \to \infty$.)*

b. *For other utility functions $U$, $\lim_{x \to \infty} u'(x) > 0$. If we denote $\lim_{x \to \infty} u'(x)$ by $y_0$, then $v(y) = \infty$ for $y < y_0$, while $v(y) < \infty$ for $y > y_0$. As for the behavior of $v$ as $y \searrow y_0$, we have the following possibilities:*

i. $\lim_{y \searrow y_0} v(y) = \infty$

ii. $\lim_{y \searrow y_0} v(y) < \infty$ *and* $\lim_{y \searrow y_0} v'(y) = \infty$;

iii. $\lim_{y \searrow y_0} v(y) < \infty$ *and* $\lim_{y \searrow y_0} v'(y) < \infty$.

*Moreover, all are possible for any value of $y_0 > 0$.*

*Finally, $AE(u) \leq AE(U)$; hence, $AE(U) < 1$ implies $\lim_{x \to \infty} u'(x) = 0$. That is, asymptotic elasticity less than 1 removes the cases given by part b.*

The possibility outlined in part *a* is simple to show: Take utility functions with constant relative risk aversion, for which solutions are well known and fit case *a*. And the final assertion needs no proof; it derives from [16], Theorem 2.2. To give examples of the three possibilities outlined in part *b* requires some calculations. Since this is a diversion from our main message, we leave this to Section 10. In fact, while part *b* seems to be the most intriguing aspect of the proposition, we note that for the proof of Theorem 1 below we only rely on the final assertion of Proposition 1.

## 5 Discrete time is asymptotically no worse than continuous time

**Proposition 2.** *For the sequence of discrete-time economies as described in Section 1, and a utility function $U$ that satisfies conditions (3.1),*

$$\liminf_{n \to \infty} u_n(x) \geq u(x) \text{ for all } x > 0.$$

Note: If $u(x) = \infty$, this proposition still applies, implying that $\lim_{n \to \infty} u_n(x) = \infty$.

*Proof.* Kreps ([17], Proposition 5.2) states that if, in the BSM economy, a bounded and continuous contingent claim $X$ satisfies $\mathbb{E}_{\mathcal{P}}[U(X)] = z$ and $\mathbb{E}_{\mathcal{P}^*}[X] = x$ (so that $u(x) \geq z$), then for every $\epsilon > 0$, there exists $N$ such that, for all $n > N$, the consumer in the $n$th discrete-time economy can synthesize a claim $X_n$ for an initial investment of $x$ such that $\mathbb{E}_{\mathcal{P}_n}[U(X_n)] \geq z - \epsilon$.[3]

Suppose we know that $u(x) < \infty$ and, for the given $x$, a solution to the consumer's problem exists (that is, the sup that defines $u(x)$ is a max). We then know, since the solution is of the form $X = I(yZ)$ for some multiplier $y > 0$ (see (4.4) above), that the solution $X : \Omega \to (0, \infty)$ is a continuous function of $\omega$. By truncating the solution $X$, we get approximately $u(x)$, with what is a bounded and continuous claim. Hence, we conclude that

$$\liminf_n u_n(x) \geq u(x).$$

The cases where $u(x) < \infty$ but no solution exists and where $u(x) = \infty$ are a bit more delicate, because we don't know, a priori, that we approach the upper bound (finite in the first case, $\infty$ in the second) with bounded and continuous contingent claims. But we can show this is so. Suppose for some level $z$, there is a measurable contingent claim $X$ such that $\mathbb{E}_{\mathcal{P}}[U(X)] = z$ and $\mathbb{E}_{\mathcal{P}^*}[X] = x$. In this context, of course $X \geq 0$.

Fix $\epsilon > 0$. We first replace $X$ with a bounded claim $X'$, bounded away from $\infty$ above and away from 0 below, in two steps. First, for $\alpha < 1$ but close to 1, let $X^\alpha := \alpha X + (1-\alpha)x$. Of course, $\mathbb{E}_{\mathcal{P}^*}[X^\alpha] = x$. And by a double application of monotone convergence (split $\mathbb{E}_{\mathcal{P}}[U(X^\alpha)]$ into $\mathbb{E}_{\mathcal{P}}[U(X^\alpha)\mathbb{1}_{\{X^\alpha \geq x\}}] + \mathbb{E}_{\mathcal{P}}[U(X^\alpha)\mathbb{1}_{\{X^\alpha < x\}}]$), we have $\lim_{\alpha \to 1} \mathbb{E}_{\mathcal{P}}[U(X^\alpha)] = \mathbb{E}_{\mathcal{P}}[U(X)] = z$. So choose $\alpha^o$ close enough to 1 so that $\mathbb{E}_{\mathcal{P}}[U(X^\alpha)] \geq z - \epsilon/4$. Of course, $X^\alpha$ is bounded below by $(1-\alpha)x$. As for the upper bound, cap $X^{\alpha^o}$ at some large $\beta$. That is, let $X^{\alpha^o, \beta}$ be $X^{\alpha^o} \wedge \beta$. For large enough $\beta^o$, this is bounded above and will satisfy $\mathbb{E}_{\mathcal{P}}[U(X^{\alpha^o, \beta^o}] > z - \epsilon/2$, while capping $X^{\alpha^o}$ can only relax the budget constraint.

So, it is wlog to assume that our original $X$ (that gives expected utility close to $z$ and satisfies the budget constraint for $x$) is bounded above and bounded away from zero. Now apply a combination of Luzin's Theorem and Tietze's Extension Theorem: We can approximate $X$ with a continuous function $X'$ that differs from $X$ on a set of arbitrarily small measure and that satsifies the same upper and lower bounds as $X$; this allows the choice of $X'$ to satisfy $\mathbb{E}_{\mathcal{P}}[U(X')] > z - 3\epsilon/4$. It may be that $\mathbb{E}_{\mathcal{P}^*}[X'] > x$, but the last $\epsilon/4$ is used to replace $X'$ with $X' - (\mathbb{E}_{\mathcal{P}^*}[X'] - x)$, giving a bounded and continuous contingent claim that costs $x$ (or less) and provides expected utility $z - \epsilon$, at which point [17], Proposition 5.2 can be applied to prove (in general) Proposition 2. □

## 6 The "relaxed" problem

In order to tackle the reverse inequality of the one in Proposition 2 we need some preparation.

---

[3]The proof of this proposition relies on Theorem 1 in [18], which says that any bounded and continuous contingent claim $x$ can be synthesized with "$x$-controlled risk" in the $n$th discrete-time economy for large enough $n$, where "approximately synthesized" means: For given $\epsilon > 0$ and large enough $n$ (depending on $\epsilon$), the synthesized claim, $x_n$ satisfies $\mathcal{P}_n(|x_n - x| > \epsilon) < \epsilon$; and "$x$-controlled risk" means that the synthesized claim $x_n$ satisfies $x_n(\omega) \in \big(\inf_{\omega'} x(\omega'), \sup_{\omega'} x(\omega')\big)$ with $\mathcal{P}_n$-probability 1.

For the $n$th discrete-time economy, the consumer faces three types of constraints:

1. She has a level of initial wealth $x$, and her initial portfolio cannot exceed $x$ in value.

2. Between times 0 and 1, any trades she makes must be self-financing.

3. She has available only the trades that the price process permits. In a word, her final consumption bundle must be a *synthesizable* contingent claim.

The importance of 3 is that, for $\zeta$ having support with more than two elements, the consumer does not face "complete markets".

However, we know that any final consumption bundle $X$ that she constructs in the n'th economy subject to these three constraints must satisfy

$$\mathbb{E}_{Q_n^*}[X] = x,$$

where $\mathbb{E}_{Q_n^*}$ denotes expectation with respect to *any* probability measure $Q_n^*$ that is an equivalent martingale measure (emm) for $\mathcal{P}_n$.[4]

We fix one particular emm for each $\mathcal{P}_n$, namely the emm, which we hereafter denote by $\mathcal{P}_n^*$, provided by the Esscher transform:

$$\left(\frac{d\mathcal{P}_n^*}{d\mathcal{P}_n}\right)(\omega) = \exp\bigl[-a_n\omega(1) - b_n\bigr]$$

for constants $a_n$ and $b_n$, chosen such that $\mathcal{P}_n^*$ is a martingale probability measure. Specifically, $a_n$ is fixed by the "martingale equation" that

$$\mathbb{E}_{\mathcal{P}_n}\left[\frac{d\mathcal{P}_n^*}{d\mathcal{P}_n}e^{\omega((k+1)/n)}\bigg|F_{k/n}\right] = e^{\omega(k/n)},$$

and $b_n$ is then fixed as a normalizing constant, given the value of $a_n$. Moreover, it can be shown that $a_n = 1/2 + \frac{E[\zeta^3]}{24\sqrt{n}} + o(1/\sqrt{n})$ where $E[\zeta^3]$ is the third moment of $\zeta$, and that $\lim_n b_n = 1/8$. (The notation $E[\cdot]$ is used to denote expectations over $\zeta$.) Of course, $\mathcal{P}_n \Rightarrow \mathcal{P}$ (weakly on $C_0[0,1]$ endowed with the sup-norm topology) and, for this specific equivalent martingale measure, $\mathcal{P}_n^* \Rightarrow \mathcal{P}^*$.[5]

So, suppose we pose the following problem for the consumer:

$$\text{Maximize } \mathbb{E}_{\mathcal{P}_n}[U(X)], \text{ subject to } \mathbb{E}_{\mathcal{P}_n^*}[X] = x, \qquad (6.1)$$

where $\mathbb{E}_{\mathcal{P}_n}[\cdot]$ denotes expectation with respect to $\mathcal{P}_n$ and $\mathbb{E}_{\mathcal{P}_n^*}[\cdot]$ denotes expectation with respect to the specific emm $\mathcal{P}_n^*$. In words, we allow the consumer any consumption claim $X$ she wishes to purchase, subject only to the constraint that she can afford $X$ at the "prices" given by $d\mathcal{P}_n^*/d\mathcal{P}_n$.

Let $Z_n$ be the function on $C_0[0,1]$ given by $Z_n(\omega) = \exp\bigl[-a_n\omega(1) - b_n\bigr]$. Hence, $Z_n$ is a specific version of the random variable $d\mathcal{P}_n^*/d\mathcal{P}_n$ and the constraint $\mathbb{E}_{\mathcal{P}_n^*}[X] = x$ can be rewritten as $\mathbb{E}_{\mathcal{P}_n}[Z_n X] = x$. That is, we can think simply of a consumer facing complete markets with $Z_n$ the "pricing kernel" for contingent claims. Using this interpretation, we denote the supremal utility the consumer can obtain in the problem (4.1) as

$$u_n^{Z_n}(x) := \sup \{\mathbb{E}_{\mathcal{P}_n}[U(X)], \text{ subject to } \mathbb{E}_{\mathcal{P}_n}[Z_n X] \leq x\}. \tag{6.2}$$

---

[4]Because we assume that $\zeta$ has mean zero and variance one, we know that $\mathcal{P}^n$ admits emms.

[5]For a detailed derivation, see[17], Lemma 5.1.

The point of this is that the problem (6.1) relaxes the constraints that actually face the consumer in the $n$th discrete-time economy; in (6.1) she faces "complete markets"; in her real problem, she faces further "synthesizability" constraints. Hence, we know that

$$u_n^{Z_n}(x) \geq u^n(x) \text{ for all } x > 0 \text{ and } n = 1, 2, \ldots \tag{6.3}$$

If we can show that $\lim_n u_n^{Z_n}(x) = u(x)$, we will know that $\limsup_n u_n(x) \leq u(x)$. This, together with Proposition 2, will establish that $\lim_n u_n(x) = u(x)$. So this is what we set out to do.

## 7   An analogous problem

In fact, we add one more plot element. As we have stated above, $Z$ is the function $Z(\omega) = \exp(-\omega(1)/2 - 1/8)$, which is a version (the unique continuous version) of $d\mathcal{P}^*/d\mathcal{P}$. Define

$$u_n^Z(x) = \sup \{\mathbb{E}_{\mathcal{P}_n}[U(X)] : \mathbb{E}_{\mathcal{P}_n}[ZX] \leq x\}. \tag{7.1}$$

In words, $u_n^Z(x)$ is the supremal expected utility that the consumer can attain if she faces complete markets and "prices" $Z$ in the $n$th discrete-time economy. That is, moving from the consumer's problem in the BSM model to the problem described by (7.1) changes the consumer's probability assessment from $\mathcal{P}$ to $\mathcal{P}_n$ but not the "prices" she faces. In moving from (7.1) to (6.2), we keep the probability assessment as $\mathcal{P}_n$ but change the prices from $Z$ to $Z_n$. This "taking it one step at a time" is useful in the analysis to follow.

## 8   If asymptotic elasticity is less than 1, optimal expected utilities are finite and converge

In this section, we prove the following result:

**Theorem 1.** *Suppose that the utility function $U$ satisfies conditions (3.1) and that $AE(U) < 1$. Then, for all $x > 0$, the value function $x \to u(x)$ is finite-valued and*

$$\lim_{n \to \infty} u_n(x) = u(x). \tag{8.1}$$

The proof of Theorem 1 will take several steps and consumes this entire section. We begin with a lemma.

**Lemma 1.** *For any constant $\gamma$, $\lim_{n \to \infty} \mathbb{E}_{\mathcal{P}_n}[\exp(\gamma \omega(1)] = \mathbb{E}_{\mathcal{P}}[\exp(\gamma \omega(1))]$.*

**Proof of Lemma 1.** Under $\mathcal{P}_n$, $\omega(1) = \sum_{k=1}^n \zeta_k/\sqrt{n}$ for $\{\zeta_k\}$ an i.i.d. sequence of random variables with the law of $\zeta$. Hence,

$$\mathbb{E}_{\mathcal{P}_n}[\exp(\gamma\omega(1))] = \mathbb{E}_{\mathcal{P}_n}\left[\exp\left(\gamma \sum_{k=1}^n \frac{\zeta_k}{\sqrt{n}}\right)\right] = \left(E\left[\exp\left(\gamma \frac{\zeta}{\sqrt{n}}\right)\right]\right)^n.$$

A Taylor series approximation to $\exp(\gamma\zeta/\sqrt{n})$ is $1+\gamma\zeta/(\sqrt{n})+\gamma^2\zeta^2/(2n)+o(1/n)$, where the $o(1/n)$ term is uniform in the value of $\zeta$ because $\zeta$ has bounded support. Therefore,

$$\mathbb{E}_{\mathcal{P}_n}\big[\exp(\gamma\omega(1))\big] = \left(E\left[1+\frac{\gamma\zeta}{\sqrt{n}}+\frac{\gamma^2\zeta^2}{2n}+o(1/n)\right]\right)^n = \big(1+\gamma^2/(2n)+o(1/n)\big)^n.$$

The term on the rhs converges to $e^{\gamma^2/2}$, which is (of course) $\mathbb{E}_{\mathcal{P}}\big[\exp(\gamma\omega(1))\big]$. □

Now we turn to the proof of Theorem 1:

**Step 1.** *Because $AE(U) < 1$, $u(x) < \infty$ for all $x > 0$.*

This step rates a remark: For "general" price processes as investigated, for instance, in [16], having asymptotic elasticity less than 1 does *not* guarantee that the optimal expected utility is finite. The result here strongly depends on the price processes being given by the BSM model.

The second key to Step 1 is the following bound:

There exist $L > 0$ and $\alpha > 0$ such that $V(y) \leq Ly^{-\alpha}$, for all $y \in (0,\infty)$. (8.2)

Corollary 6.1 in [16] establishes this bound as a consequence of $AE(U) < 1$, but only for $0 < y \leq y_0$, for some $y_0 > 0$. If we have that $V(\infty) = U(0) < 0$, we get the bound for all $y > 0$. And, for purposes of this theorem, it is without loss of generality to shift $U$ by a constant. So if $U(0) \geq 0$, simply replace $U$ with $U(x) - U(0) - c$, for a suitable constant $c > 0$. Then we have, and henceforth assume, 8.2 for all $y \in (0,\infty)$.

Hence, we may estimate

$$v(y) = \mathbb{E}_{\mathcal{P}}\left[V\left(y\frac{dP^*}{dP}\right)\right] \leq L\,\mathbb{E}_{\mathcal{P}}\left[y^{-\alpha}\exp\left(-\frac{\omega(1)}{2}-\frac{1}{8}\right)^{-\alpha}\right]$$
(8.3)
$$= L\,y^{-\alpha}e^{\alpha/8}\,\mathbb{E}\left[\exp\left(\frac{\alpha}{2}\omega(1)\right)\right] < \infty,$$

as the latter expectation is just an exponential moment of a Gaussian variable. Hence, by ([16], Theorem 2.0), the dual value function $y \to v(y)$ as well as the primal value function $x \to u(x)$ have finite values; and we deduce as well from part $c$ of Proposition 1, that $AE(u) \leq AE(U) < 1$.

**Step 2.** *Using the notation (7.1), define $u_\infty^Z(x)$ for $x > 0$ by*

$$u_\infty^Z(x) := \limsup_{n\to\infty} u_n^Z(x). \tag{8.4}$$

*Then $u_\infty^Z(x) < \infty$ for all $x > 0$.*

Define for each $n$ the conjugate function $v_n^Z$, which is

$$v_n^Z(y) = \mathbb{E}_{\mathcal{P}_n}[V(yZ)]. \tag{8.5}$$

By the same argument that gave (8.3), we have

$$v_n^Z(y) = \mathbb{E}_{\mathcal{P}_n}\left[V(yZ)\right] \le L\, y^{-\alpha}\, e^{\alpha/8}\, \mathbb{E}_{\mathcal{P}_n}\left[\exp\left(\frac{\alpha}{2}\omega(1)\right)\right]. \tag{8.6}$$

Lemma 1 tells us that the expectation on the rhs of (8.6) converges, which implies that $v_n^Z(y)$ is uniformly bounded in $n$ for fixed $y$, which, by standard arguments concerning conjugate functions, proves that $u_\infty^Z(x)$ is finite for each $x$.

**Step 3.** *Indeed, $\lim_n u_n^Z(x)$ exists and equals $u(x)$ for all $x > 0$.*

We show that $\lim_n v_n^Z(y) = v(y)$ for all $y > 0$, which proves step 3, again using standard arguments concerning conjugate functions.

Compare $v_n^Z(y)$ and $v(y)$:

$$v_n^Z(y) = \mathbb{E}_{\mathcal{P}_n}[V(yZ)] \quad \text{and} \quad v(y) = \mathbb{E}_{\mathcal{P}}[V(yZ)].$$

If $V$ were a bounded function (of course, $V$ is continuous), the conclusion would follow immediately from $\mathcal{P}_n \Rightarrow \mathcal{P}$. But $V$ is typically not bounded, and so we must show that the contributions to the expectations from the "tails" can be uniformly controlled. We do this by showing the following two uniform bounds:

For every $y > 0$ and $\epsilon > 0$, there exists $M > 0$ such that
$$\mathbb{E}_{\mathcal{P}_n}\left[|V(yZ)| \cdot \mathbb{1}_{\{V(yZ) < -M\}}\right] < \epsilon, \text{ uniformly in } n. \tag{8.7}$$

For every $y > 0$ and $\epsilon > 0$, there exists $M > 0$ such that
$$\mathbb{E}_{\mathcal{P}_n}\left[V(yZ) \cdot \mathbb{1}_{\{V(yZ) > M\}}\right] < \epsilon, \quad \text{uniformly in } n. \tag{8.8}$$

Begin with (8.7). If $U(0)$ is finite, it follows that $V(y) \ge V(\infty) = U(0)$ for all $y$, so taking $M = -U(0)$ immediately works. The (slightly) harder case is where $U(0) = -\infty$. In this case, recall that, by the Inada conditions, $\lim_{y \to \infty} V'(y) = -\lim_{y \to \infty}(U')^{-1}(y) = 0$. Since $V$ is convex, this implies that for large enough $M$, $|V(y)| \le \epsilon y$, provided $V(y) \le -M$. But then

$$\mathbb{E}_{\mathcal{P}_n}\left[|V(yZ)| \cdot \mathbb{1}_{\{V(yZ) \le -M\}}\right] \le \mathbb{E}_{\mathcal{P}_n}\left[\epsilon Z\right] \le \epsilon \mathbb{E}_{\mathcal{P}_n}\left[Z\right]. \tag{8.9}$$

By Lemma 1, $\lim_{n \to \infty} \mathbb{E}_{\mathcal{P}_n}[Z] = \mathbb{E}_{\mathcal{P}}[Z] = 1$, which shows (8.7).

And to show the (uniform) inequality (8.8): For the parameters $\alpha$ and $L$ that give the bound (8.2) and for fixed $y > 0$ and $\epsilon > 0$, let $B$ be large enough so that

$$\mathbb{E}_{\mathcal{P}_n}\left[e^{\alpha \omega(1)/2} \cdot \mathbb{1}_{\{\omega(1) \ge B\}}\right] \le \frac{\epsilon}{L y^{-\alpha} e^{\alpha/8}} \quad \text{for all } n. \tag{8.10}$$

The existence of such a $B$ follows from Lemma 1 and, because $\mathcal{P}_n \Rightarrow \mathcal{P}$,
(1) $\mathbb{E}_{\mathcal{P}_n}\left[\min\{e^{\alpha\omega(1)}, B\}\right] \to \mathbb{E}_{\mathcal{P}}\left[\min\{e^{\alpha\omega(1)}, B\}\right]$ and
(2) $\mathcal{P}_n(\{\omega(1) \ge B\}) \to \mathcal{P}(\{\omega(1) \ge B\})$, for all $B > 0$.

And let
$$M = Ly^{-\alpha} \exp\left(\frac{\alpha B}{2} + \frac{\alpha}{8}\right). \tag{8.11}$$

We have $V(y) \leq Ly^{-\alpha}$ for all $y > 0$, and so

$$\{V(yZ) \geq M\} \subseteq \{L(yZ)^{-\alpha} \geq M\} = \{L(yZ)^{-\alpha} \geq Ly^{-\alpha} e^{(\alpha B/2 + \alpha/8)}\}$$
$$= \{Z^{-\alpha} \geq e^{\alpha B/2 + \alpha/8}\} = \{(e^{-\omega(1)/2 - 1/8})^{-\alpha} \geq e^{\alpha B/2 + \alpha/8}\}$$
$$= \{e^{\alpha\omega(1)/2} \geq e^{\alpha B/2}\} = \{\omega(1) \geq B\}.$$

Hence,

$$\mathbb{E}_{\mathcal{P}_n}\left[V(yZ) \cdot \mathbb{1}_{\{V(yZ) \geq M\}}\right] \leq \mathbb{E}_{\mathcal{P}_n}\left[L(yZ)^{-\alpha} \cdot \mathbb{1}_{\{\omega(1) \geq B\}}\right]$$
$$= \mathbb{E}_{\mathcal{P}_n}\left[Ly^{-\alpha}(e^{-\omega(1)/2-1/8})^{-\alpha} \cdot \mathbb{1}_{\{\omega(1) \geq B\}}\right]$$
$$= Ly^{-\alpha} e^{\alpha/8} \mathbb{E}_{\mathcal{P}_n}[e^{\alpha\omega(1)/2} \cdot \mathbb{1}_{\{\omega(1) \geq B\}}] \leq \epsilon,$$

uniformly in $n$.

Having shown the two uniform bounds (8.7) and (8.8), what remains is a standard argument. Fix $y$ and $\epsilon$ both $> 0$, and find $M$ such that (8.7) and (8.8) both hold. Let $V^M(yZ) := \max\{-M, \min\{M, V(yZ)\}\}$; that is, $V^M(yZ)$ is $V(yZ)$ "truncated" at $\pm M$. This truncated function is bounded and continuous, so $\mathcal{P}_n \Rightarrow \mathcal{P}$ implies $\mathbb{E}_{\mathcal{P}_n}[V^M] \to \mathbb{E}_{\mathcal{P}}[V^M]$. And the differences $\mathbb{E}_{\mathcal{P}_n}[V(yZ) - V^M(yZ)]$ are uniformly bounded by $2\epsilon$. Hence, $v_n(y) = \mathbb{E}_{\mathcal{P}_n}[V(yZ)] \to \mathbb{E}_{\mathcal{P}}[V(yZ)] = v(y)$ for all $y$, which implies that $u_n^Z(x) \to u(x)$ for all $x > 0$.

**Step 4.** *For every $y > 0$ and $\epsilon > 0$, there exists $M > 0$ such that*

$$\mathbb{E}_{\mathcal{P}_n}\left[|V(yZ^n)| \cdot \mathbb{1}_{\{V(yZ^n) < -M\}}\right] < \epsilon, \quad \text{uniformly in } n. \tag{8.12}$$

The parallel to the uniform inequality (8.7) is obvious: (8.7) uniformly controls the right-hand tail of the integral $v_n^Z(y) = \mathbb{E}_{\mathcal{P}_n}[V(yZ)]$; here we are uniformly controlling the right-hand tail of the integral $v_n^{Z_n}(y) = \mathbb{E}_{\mathcal{P}_n}[V(yZ_n)]$. And the same proof works; indeed, in this case we even have $\mathbb{E}_{\mathcal{P}_n}[Z_n] = 1$ (as opposed to $\mathbb{E}_{\mathcal{P}_n}[Z] \to 1$ before).

**Step 5.** *There is a constant $C > 1$ such that, for all $n$,*

$$\frac{1}{C} \leq \frac{Z_n(\omega)}{Z(\omega)} \leq C, \mathcal{P}_n - a.s. \tag{8.13}$$

(The reason for this step is to prove a uniform bound for $\mathbb{E}_{\mathcal{P}_n}[V(yZ_n)]$ analogous to (8.8), which is Step 6.)

We have that

$$\frac{Z_n(\omega)}{Z(\omega)} = \frac{e^{-a_n\omega(1) - b_n}}{e^{-\omega(1)/2 - 1/8}},$$

where $a_n = 1/2 + d/\sqrt{n} + o(1/\sqrt{n})$, where $d = E[\zeta^3]/24$, and $b_n = 1/8 + o(1)$. Hence,

$$\frac{Z_n(\omega)}{Z(\omega)} = \exp\left[\frac{d\,\omega(1)}{\sqrt{n}} + \omega(1) \cdot o\left(\frac{1}{\sqrt{n}}\right) + o(1)\right].$$

Since $\zeta$ has bounded support, there is some constant $K$ such that $|\zeta| \leq K$ with probability 1, and so $|\omega(1)| \leq K\sqrt{n}$, $\mathcal{P}_n$-a.s. The ability to find a constant $C > 1$ that gives (8.13) is now evident.

**Step 6.** *For every $y > 0$ and $\epsilon > 0$, there exists $M > 0$ such that*
$$\mathbb{E}_{\mathcal{P}_n}\big[V(yZ_n) \cdot \mathbb{1}_{\{V(yZ_n) > M\}}\big] < \epsilon, \quad \text{uniformly in } n. \tag{8.14}$$

Rewrite the left-hand inequality in (8.13) as $Z(\omega)/C \leq Z_n(\omega)$, on the support of $\mathcal{P}_n$. Since $V$ is a decreasing function, this implies that, for all $y > 0$,

$$V(yZ(\omega)/C) \geq V(yZ_n(\omega)) \quad \text{and, therefore,}$$
$$\{V(yZ(\omega)/C) > M\} \supseteq \{V(yZ_n(\omega)) > M\},$$

both restricted to the support of $\mathcal{P}_n$. Therefore, for any $M > 0$,

$$\mathbb{E}_{\mathcal{P}_n}\big[V(yZ_n) \cdot \mathbb{1}_{\{V(yZ_n) > M\}}\big] \leq \mathbb{E}_{\mathcal{P}_n}\big[V(yZ(\omega)/C) \cdot \mathbb{1}_{\{V(yZ(\omega)/C) > M\}}\big].$$

But then the proof of the existence of $M$ such that (8.8) is satisfied can be applied to $y' = y/C$, which completes this step.

**Step 7.** *For all $y > 0$,*

$$\lim_n \big|\mathbb{E}_{\mathcal{P}_n}[V(yZ_n)] - \mathbb{E}_{\mathcal{P}_n}[V(yZ)]\big| = 0.$$

The argument for this step changes a bit when $V(0) = U(\infty)$ and/or $V(\infty) = U(0)$ are finite valued. So we first give the argument in the case where $V(0) = U(\infty) = \infty$ and $V(\infty) = U(0) = -\infty$, and then sketch how to handle the easier cases where one or the other is finite.

Fix $\epsilon > 0$ and $y > 0$, and pick $M > 0$ large enough so that (8.7), (8.8), (8.12), and (8.14) all hold. Let $M' = M + 2$. Let $w_1$ and $w_2$ be the solutions, respectively, to $V(yZ(w_1)) = M'$ and $V(yZ(w_2)) = -M'$, where by $Z(w)$, we temporarily mean $\exp(-w/2 - 1/8)$. This implies that if $\omega$ is such that $\omega(1) \in [w_1, w_2]$, then $V(yZ(\omega)) \in [-M', M']$. Moreover, by the continuity and monotonicity of $V$, $Z$, and $Z_n$ (the latter two viewed as functions of $\omega(1)$), and the fact that for fixed $w$, $Z_n(w) \to Z(w)$, there exists $n_0$ such that for all $n > n_0$, $\omega(1) \in [w_1, w_2]$ implies that $V(yZ_n(\omega)) \in [-M - 1, M + 1]$.

The functions $V$, $Z$, and $Z_n$ are all uniformly and even Lipschitz continuous on compact domains (for $V$, that are strictly bounded away from 0), so there is a Lipschitz constant $\ell$ such that
$$\big|V(yZ_n(w)) - V(yZ(w))\big| \leq \ell \cdot \big|Z_n(w) - Z(w)\big|,$$
if $n > n_0$ and $w \in [w_1, w_2]$. And, on the event $\omega(1) \in [w_1, w_2]$, there is $n_1$ such that for all $n > n_1$, $|Z_n(\omega) - Z(\omega)| < \epsilon/\ell$.

Hence, on the event $\mathcal{D} = \{\omega(1) \in [w_1, w_2]\}$, and for $n > \max\{n_0, n_1\}$, we have $|V(yZ(\omega)) - V(yZ_n(\omega))| \leq \epsilon$, and $\mathbb{E}_{\mathcal{P}_n}\big[|V(yZ(\omega)) - V(yZ_n(\omega))| \cdot \mathbb{1}_{\mathcal{D}}\big] < \epsilon$. By construction, the complement of $\mathcal{D}$ is a subset of the union of the four events on which we have uniformly controlled the integrals of $V(yZ(\omega))$ and $V(yZ_n(\omega))$, so for $n > \max\{n_0, n_1\}$, $\mathbb{E}_{\mathcal{P}_n}\big[|V(yZ(\omega)) - V(yZ_n(\omega))|\big] < 5\epsilon$, uniformly in $n$, which proves Step 7.

When $V(0)$ and/or $V(\infty)$ are finite, the argument needs a bit of modification. Suppose $V(0) < \infty$. This is relevant when $yZ$ and $yZ_n$ are both close to zero, which is for paths $\omega$ where $\omega(1)$ is large. And for those paths, $Z_n(\omega)$ can be quite far from $Z(\omega)$. However even if these terms are far apart, $V(yZ(\omega))$ and $V(yZ_n(\omega))$ will be close together, since each is close to the finite $V(0)$. A similar argument works for cases where $V(\infty)$ is finite.

**Step 8.** *Combine Steps 7 and 3 to conclude that* $\lim_{n\to\infty} u_n^{Z_n} = u(z)$ *for all* $z > 0$.

Step 3 shows that $v_n^Z(y) \to v(y)$ for all $y > 0$ (which is how we concluded that $u_n^Z(x) \to u(x)$). Step 7 then implies that $v_n^{Z_n}(y) \to v(y)$ for all $y > 0$. This, in turn, implies $\lim_{n\to\infty} u_n^{Z_n}(x) \to u(x)$ by standard arguments on conjugate functions.

**Step 9.** *Combine Steps 8 and Proposition 2 to finish the proof.*
The argument has already been given. □

This proof clarifies why we introduced the analogous problem, where a consumer with probability assessment $\mathcal{P}_n$ faces complete markets and prices given by $Z$: Comparing this with the BSM model, the conjugates $v_n$ and $v$ to optimal expected utility functions $u_n$ and $u$ are the expectations of a fixed function for different probability measures. So, after controlling the tails of the integrals that define these conjugate functions, we have a more or less standard consequence-of-weak-convergence result in Step 3. In Step 7, *both* the probability assessments and the prices (for one of the two problems being compared) change with $n$. While the pairs of problems being compared differ only in the prices, because *both* the integrand and the integrating measure $\mathcal{P}_n$ change with $n$, a level of finicky care is required.

# 9   A counterexample to Kreps' conjecture

Theorem 1 guarantees that for utility functions $U$ that satisfy the conditions (3.1) and have asymptotic elasticity less than one, everything works out nicely *within the context of the BSM model and the discrete-time approximations to BSM that we have posited*.

It is natural to ask, then, what can be said if we maintain (3.1) and these specific models of the financial markets, but we look at utility functions $U$ for which $\mathrm{AE}(U) = 1$. In such cases, it may be that things work out in the sense of Theorem 1. But it is also possible that $\limsup_{n\to\infty} u_n(x) > u(x)$. That is, when $\mathrm{AE}(U) = 1$, Kreps' conjecture can fail. In this section, we provide an example to illustrate this failure in stark fashion: In this example, $u(x) < \infty$ while $\limsup_{n\to\infty} u_n(x) = \infty$, both for all $x > 0$.

In this example (and also in Section 9, where we finish the proof of Proposition 1), we construct conjugate functions $V$ taking the form

$$V(y) := \sum_{k=1}^{\infty} \beta_k y^{-\alpha_k},$$

where $\alpha_k, \beta_k > 0$ and the sequences $\{\alpha_k\}$ and $\{\beta_k\}$ are chosen so that the sum defining $V(y)$ is finite for all $y > 0$.

We begin with some standard facts about conjugate pairs $U$ and $V$, when $V$ has the form $V(y) = \beta y^{-\alpha}$.

**Lemma 2.** *For $\alpha > 0$ and $\beta > 0$, denote by $V_{\alpha,\beta}(y)$ the function $V_{\alpha,\beta}(y) := \beta y^{-\alpha}$ for $y > 0$. For the utility function $U_{\alpha,\beta}$ that is conjugate to $V_{\alpha,\beta}$, if $x_0 = -V'_{\alpha,\beta}(y_0) = \beta\alpha y_0^{-\alpha-1}$ for given $y_0$, then*

$$U_{\alpha,\beta}(x_0) = (1+\alpha)V_{\alpha,\beta}(y_0). \tag{9.1}$$

The resulting utility function $U_{\alpha,\beta}$ is

$$U_{\alpha,\beta}(x) = \frac{1+\alpha}{\alpha^{\alpha/(1+\alpha)}} \beta^{1/(1+\alpha)} x^{\alpha/(1+\alpha)}, \quad \text{for } x > 0.$$

Since $\alpha > 0$, $\alpha/(1+\alpha) \in (0,1)$, and $AE(U_{\alpha,\beta}) = \alpha/(1+\alpha) < 1$.

Lemma 3 provides some analysis of the consumer's maximization problem in the context of the BSM model, when the conjugate to her utility function has the form $V_{\alpha,\beta}$.

**Lemma 3.** *Imagine a consumer in the BSM economy whose utility function $U_{\alpha,\beta}$ is given by (9.1). The (dual) value function corresponding to $U_{\alpha,\beta}$ and $V_{\alpha,\beta}$ in the BSM economy is*

$$v_{\alpha,\beta}(y) = \mathbb{E}_{\mathcal{P}}\left[V_{\alpha,\beta}(yZ)\right] = \beta e^{(\alpha^2+\alpha)/8} y^{-\alpha} = e^{(\alpha^2+\alpha)/8} V_{\alpha,\beta}(y). \tag{9.2}$$

*And the primal expected-utility function, giving the supremal expected utility that the consumer can achieve in the BSM economy as a function of her initial wealth $x$, is*

$$u_{\alpha,\beta}(x) = e^{\alpha/8} U_{\alpha,\beta}(x). \tag{9.3}$$

*Proof.* Equation (9.3) is easily derived from (9.2), so we only give the proof of (9.2).

Let $Y$ be $N(0,1)$-distributed so that $Y^{-1/8, 1/4} = -Y/2 - 1/8$ has the law of $\ln(d\mathcal{P}^*/d\mathcal{P}) = \ln(Z)$. Hence, the random variable $\beta\left[y \exp(-Y/2 - 1/8)\right]^{-\alpha}$ has the law of $V_{\alpha,\beta}(yZ)$, and so

$$\begin{aligned} v_{\alpha,\beta}(y) &= \mathbb{E}_{\mathcal{P}}\left[\beta\left(y \exp\left(\frac{-Y}{2} - \frac{1}{8}\right)\right)^{-\alpha}\right] \\ &= \beta y^{-\alpha} e^{\alpha/8} \mathbb{E}_{\mathcal{P}}\left[\exp\left(\frac{\alpha Y}{2}\right)\right] = \beta e^{(\alpha+\alpha^2)/8} y^{-\alpha}. \end{aligned}$$

$\square$

The factor $e^{(\alpha+\alpha^2)/8}$ recurs occasionally, so to save on keystrokes, let $\phi(\alpha) := e^{(\alpha+\alpha^2)/8}$.

Denote by $\mathcal{L}_\zeta(\lambda)$ the Laplace transform of the law of $\zeta$; that is

$$\mathcal{L}_\zeta(\lambda) := E[\exp(\lambda \zeta)].$$

As above, denote by $Y$ a standard (mean 0, variance 1) Normal variate and write

$$\mathcal{L}_Y(\lambda) := E[\exp(\lambda Y)] = e^{\lambda^2/2}.$$

Letting $Y_n$ be the scaled sum of $n$ independent copies of $\zeta$,

$$Y_n = \frac{\zeta_1 + \ldots + \zeta_n}{n^{1/2}},$$

we have

$$\mathcal{L}_{Y_n}(\lambda) = \mathcal{L}_\zeta\left(\frac{\lambda}{n^{1/2}}\right)^n.$$

The Central Limit Theorem corresponds to the well known fact that $\mathcal{L}_{Y_n}(\lambda)$ converges to $\mathcal{L}_Y(\lambda)$. On the other hand, if $E[\zeta^3] > 0$, by considering — similarly as in the proof of Lemma 1 — the Taylor series expansion up to degree 3 of $\exp(\lambda Y)$ around $\lambda = 0$, it follows that, for small enough $\lambda_0 > 0$,

$$\mathcal{L}_\zeta(\lambda_0) > \mathcal{L}_Y(\lambda_0). \tag{9.4}$$

Now consider a conjugate utility function $V_{\alpha,\beta}(y) = \beta y^{-\alpha}$ as above, its conjugate $U_{\alpha,\beta}$, the corresponding value functions for the BSM economy $u_{\alpha,\beta}$ and its conjugate $v_{\alpha,\beta}$, as in Lemma 3, which we compare to the value functions for the various discrete-time economies where the consumer faces complete markets and prices given by $Z_n$. In this section, we do not require value functions for discrete-time economies in which the consumer faces prices $Z$, so to simplify notation, in this section we write $v_{\alpha,\beta}^n$ for the conjugate-to-the-value-function for the $n$th discrete-time economy — that is,

$$v_{\alpha,\beta}^n(y) := \mathbb{E}_{\mathcal{P}_n}\big[\beta(yZ_n)^{-\alpha}\big] \quad \text{for all } y > 0. \tag{9.5}$$

And we write $u_{\alpha,\beta}^n$ to denote that primal value function (the conjugate to $v_{\alpha,\beta}^n$).

Consider, for integer $k$, the ratio $v_{\alpha,\beta}^n(k)/v_{\alpha,\beta}(1/k)$. From (9.2) and (9.5), it is evident that this ratio is independent of the value of $\beta$, and so, for integers $k$ and $n$, and $\alpha > 0$ (and any $\beta > 0$), let

$$M(k, n, \alpha) := \frac{v_{\alpha,\beta}^n(k)}{v_{\alpha,\beta}(1/k)}. \tag{9.6}$$

**Lemma 4.** *For each integer $k$, there exists $n$ large enough so that, for $\alpha := 2\lambda_0 n^{1/2}$, $M(k, n, \alpha) \geq 2^{2k}$.*

*Proof.* Fix $k$. Without loss of generality, set $\beta = 1$. For given $n$ and $\alpha = 2\lambda_0 n^{1/2}$, calculate the denominator and numerator in the $M(k, n, \alpha)$ separately.

For the denominator, we have

$$v_{\alpha,1}\left(\frac{1}{k}\right) = \mathbb{E}_\mathcal{P}\left[\left(\frac{1}{k}\exp\left(-\frac{\omega(1)}{2} - \frac{1}{8}\right)\right)^{-\alpha}\right] = \left(\frac{1}{k}\right)^{-1/\alpha} e^{\alpha/8} \mathbb{E}_\mathcal{P}\big[\lambda_0 n^{1/2}\omega(1)\big]$$

$$= \left(\frac{1}{k}\right)^{-\alpha} e^{\alpha/8} E\big[\exp(\lambda_0 n^{1/2} Y)\big] = \left(\frac{1}{k}\right)^{-\alpha} e^{\alpha/8} \mathcal{L}_Y(\lambda_0 n^{1/2})$$

$$= \left(\frac{1}{k}\right)^{-\alpha} e^{\alpha/8} (e^{\lambda_0^2/2})^n = \left(\frac{1}{k}\right)^{-\alpha} e^{\alpha/8} \mathcal{L}_Y(\lambda_0)^n,$$

where $Y$ is a standard (mean 0, variance 1) Normal variate and $E$ denotes the expectation with respect to $Y$.

And for the numerator, which we calculate for general $y$ and $\beta$ before specializing to $y = k$ and $\beta = 1$:

$$\begin{aligned}
v_{\alpha,\beta}^n(y) &= \mathbb{E}_{\mathcal{P}_n}\big[\beta(yZ_n)^{-\alpha}\big] = \mathbb{E}_{\mathcal{P}_n}\big[\beta\big(y\exp(-a_n\omega(1) - b_n)\big)^{-\alpha}\big] \\
&= \beta y^{-\alpha} e^{\alpha b_n} E\left[\exp\left(2\lambda_0 a_n n^{1/2}\left(\frac{\zeta_1}{n^{1/2}} + \ldots + \frac{\zeta_n}{n^{1/2}}\right)\right)\right] \\
&= \beta y^{-\alpha} e^{\alpha b_n} \big(E[\exp(2a_n\lambda_0\zeta)]\big)^n = \beta y^{-\alpha} e^{\alpha b_n} \mathcal{L}_\zeta(2a_n\lambda_0)^n \\
&= y^{-\alpha} H(n, \alpha, \beta) \quad \text{for} \quad H(n, \alpha, \beta) := \beta e^{\alpha b_n} \mathcal{L}_\zeta(2a_n\lambda_0)^n,
\end{aligned} \tag{9.7}$$

where the $\zeta_j$'s are i.i.d. copies of $\zeta$, and $E$ denotes expectation with respect to these random variables.

We therefore have that

$$M(k, n, \alpha) = \frac{v_{\alpha,1}^n(k)}{v_{\alpha,1}(1/k)} = \frac{k^{-\alpha} e^{\alpha b_n} \big(\mathcal{L}_\zeta(2a_n\lambda_0)\big)^n}{(1/k)^{-\alpha} e^{\alpha/8} \big(\mathcal{L}_Y(\lambda_0)\big)^n}$$
$$= \left[k^{-4\lambda_0} e^{2\lambda_0(b_n - 1/8)}\right]^{n^{1/2}} \left[\frac{\mathcal{L}_\zeta(2a_n\lambda_0)}{\mathcal{L}_Y(\lambda_0)}\right]^n. \tag{9.8}$$

The term within the first square brackets on the rhs of (9.7) has a finite limit (since $b_n \to 1/8$), so as $n \to \infty$, this term, raised to the power $n^{1/2}$ is bounded above by $G^{n^{1/2}}$ for some constant $G$. And the term within the second set of square brackets converges to $\mathcal{L}_\zeta(\lambda_0)/\mathcal{L}_Y(\lambda_0)$, which, per (9.4), is a constant strictly greater than 1. This term is raised to the power $n$. Hence, for fixed $k$, the second term overwhelms the first term for large enough $n$, proving Lemma 4.

□

For each $k = 0, 1, 2, \ldots,$, let $n_k$ and $\alpha_k = 2\lambda_0 n_k^{1/2}$ be the values of $n$ and $\alpha$ guaranteed by Lemma 4. That is, for each $k$ (and for all $\beta > 0$),

$$M(k, n_k, \alpha_k) = \frac{v_{\alpha_k,\beta_k}^{n_k}(k)}{v_{\alpha_k,\beta_k}(1/k)} > 2^{2k}. \tag{9.9}$$

It is clear that we can add the requirements that $n_k \geq k$ and $n_k/k$ is increasing, and we do so. Since $\alpha_k = 2\lambda_0 n_k^{1/2}$, this implies that $\lim_{k \to \infty} \alpha_k = \infty$.

Let

$$\beta_k := \frac{1}{2^k v_{\alpha,1}(1/k)} \quad \text{so that} \quad \beta_k\, v_{\alpha_k,1}(1/k) = v_{\alpha_k,\beta_k}(1/k) = \frac{1}{2^k}. \tag{9.10}$$

By (9.9),

$$\frac{v_{\alpha_k,\beta_k}^{n_k}(k)}{2^k v_{\alpha_k,\beta_k}(1/k)} = v_{\alpha_k,\beta_k}^{n_k}(k) > 2^k. \tag{9.11}$$

Define

$$V(y) := \sum_{k=1}^\infty \beta_k y^{-\alpha_k} = \sum_{k=1}^\infty V_{\alpha_k,\beta_k}(y), \quad \text{for all } y > 0. \tag{9.12}$$

Clearly, the sum is well defined for all $y > 0$, and the function has all the properties required to be conjugate to a utility function $U$ that satisfies Condition (3.1).[6] Indeed, (9.10) ensures that

$$v(y) = \mathbb{E}_\mathcal{P}\big[V(yZ)\big] = \mathbb{E}_\mathcal{P}\left[\sum_{k=1}^\infty V_{\alpha_k,\beta_k}(yZ)\right] = \sum_{k=1}^\infty v_{\alpha_k,\beta_k}(y) < \infty, \quad \text{for all } y > 0,$$

which implies that, for $U$ the conjugate (utility) function to $V$, and for $u$ the (maximal expected) utility-of-wealth function corresponding to this $U$ within the BSM economy, $u(x) < \infty$ for all $x$.

---

[6] It may be worth pointing out, however, that this utility function $U$ is not $\sum_k U_{\alpha_k,\beta_k}$.

Concerning the discrete-time economies, begin by noting that all the terms in the sum (9.12) are positive, so $V_{\alpha_k,\beta_k}(y) \leq V(y)$. This implies that, for each $k$,

$$v^{n_k}_{\alpha_k,\beta_k}(y) \leq v^{n_k}(y) \text{ for all } y, \text{ and therefore } u^{n_k}_{\alpha_k,\beta_k}(x) \leq u^{n_k}(x) \text{ for all } x > 0. \qquad (9.13)$$

Now we enlist Lemma 2. Using the notation from (9.7), for arbitrary $y_k > 0$, let

$$x_k := -\left(\frac{dv^{n_k}_{\alpha_k,\beta_k}}{dy}\right)(y_k) = \alpha_k\, y_k^{-\alpha_k - 1}\, H(n_k, \alpha_k, \beta_k) = \frac{\alpha_k}{y_k} \cdot v^{n_k}_{\alpha_k,\beta_k}(y_k).$$

Lemma 2 tells us that

$$u^{n_k}_{\alpha_k,\beta_k}(x_k) = v^{n_k}_{\alpha_k,\beta_k}(y_k)(1 + \alpha_k).$$

Hence,

$$\frac{u^{n_k}_{\alpha_k,\beta_k}(x_k)}{x_k} = \frac{(1+\alpha_k)v^{n_k}_{\alpha_k,\beta_k}(y_k)}{(\alpha_k/y_k)v^{n_k}_{\alpha_k,\beta_k}(y_k)} = \frac{1+\alpha_k}{\alpha_k} y_k > y_k. \qquad (9.14)$$

Choose $y_k = k^{1/2}$. Since $v^{n_k}_{\alpha_k,\beta_k}(k) > 2^k$ and $v^{n_k}_{\alpha_k,\beta_k}(y)$ is decreasing in $y$, we know that $v^{n_k}_{\alpha_k,\beta_k}(k^{1/2}) > 2^k$. Since $\alpha_k = 2\lambda_0 n_k^{1/2}$ and $n_k/k$ is, by construction, nondecreasing, we know that $\alpha_k/k^{1/2}$ is nondecreasing. Putting these two observations together, we know that

$$\lim_{k\to\infty} x_k = \lim_{k\to\infty} \frac{\alpha_k}{k^{1/2}} \cdot v^{n_k}_{\alpha_k,\beta_k}(k^{1/2}) = \infty. \qquad (9.15)$$

And, for this choice of $y_k$, (9.14) tells us that

$$\lim_{k\to\infty} \frac{u^{n_k}_{\alpha_k,\beta_k}(x_k)}{x_k} = \lim_{k\to\infty} \frac{1+\alpha_k}{\alpha_k} k^{1/2} = \infty. \qquad (9.16)$$

Each function $u^{n_k}_{\alpha_k,\beta_k}$ is concave and has value 0 at $x = 0$. So (9.16) implies that, over the interval $x \in (0, x_k]$, $u^{n_k}_{\alpha_k,\beta_k}(x) > k^{1/2}x$. Hence, from (9.13), the same is true for $u^{n_k}(x)$. But as $k$ increases toward $\infty$, the intervals $[0, x_k]$ over which this is true expand to all of $(0, \infty]$ — this is (9.15) — and the underestimate of $u^{n_k}$ on this interval approaches infinity. This implies that

$$\lim_{k\to\infty} u^{n_k}(x) = \infty, \quad \text{for all } x > 0. \qquad (9.17)$$

The limit established in (9.17) doesn't quite accomplish what we set out to do. We want to show that, in the $n$th discrete-time economy, where the consumer faces prices $Z_n$ *and the constraint that she must be above to synthesize her consumption claim*, she can (at least, along a subsequence) asymptotically generate infinite expected utility, although she can only generate finite expected utility in the BSM economy. The limit in (9.17) concerns what expected utility she can generate facing prices $Z_n$ and complete markets.

But this final step is easy. The properties of $\zeta$ that are used to get to (9.16) (in contrast to the finiteness of supremal expected utility in the limit BSM economy) are (1) $\zeta$ has mean zero, (2) $\zeta$ has variance 1, (3) $\zeta$ has finite support, and (4) $E[\zeta^3] > 0$. For example, suppose $\zeta$ is the asymmetric binomial

$$\zeta = \begin{cases} 2, & \text{with probability } 1/5, \text{ and} \\ -1/2, & \text{with with probability } 4/5. \end{cases}$$

It is straightforward to verify that all four required properties are satisfied. And, since $\zeta$ has two-element support, for any $n$, it gives complete markets. For this asymmetric binomial $\zeta$, $u^{n_k}(x)$ is precisely what she can attain in the $n_k$th discrete-time economy, even with the synthesizability constraint imposed.

We therefore have the desired counterexample to Kreps' conjecture.

We come to the same conclusion for any asymmetric binomial with mean zero and an "uptick" greater in absolute value than the "downtick", as this gives $E[\zeta^3] > 0$.[7] It is natural to ask, then, what happens in the case of the symmetric binomial, where $\zeta = \pm 1$, each with probability $1/2$ or, more generally, for any asymmetric binomial with $E[\zeta^3] \leq 0$ or, even more generally, any $\zeta$ with mean zero, bounded support, and $E[\zeta^3] \leq 0$.

For such $\zeta$, the above reasoning does not apply. To the contrary, in the specific case of the symmetric random walk, we have

$$\mathcal{L}_\zeta(\lambda) = \cosh(\lambda) \leq e^{\lambda^2/2} = \mathcal{L}_Y(\lambda), \text{ for all } \lambda \in R. \tag{9.18}$$

This is most easily seen by comparing the Taylor series

$$\cosh(\lambda) = \sum_{k=0}^{\infty} \frac{\lambda^{2k}}{(2k)!} \quad \text{versus} \quad e^{\lambda^2/2} = \sum_{k=0}^{\infty} \frac{(\lambda^2/2)^k}{k!} = \sum_{k=0}^{\infty} \frac{\lambda^{2k}}{2^k k!}.$$

Hence, the logic of the counterexample constructed above, which requires $\mathcal{L}_\zeta(\lambda_0) > \mathcal{L}_Y(\lambda_0)$ for some $\lambda_0 > 0$, fails.

It may still be true for the symmetric binomial that $\limsup_n u_n(x) > u(x)$, for some utility function $U$ (necessarily, in view of Theorem 1, satisfying $\mathrm{AE}(U) = 1$). Or it may be that equality holds true, in the case of the symmetric binomial (and, perhaps, in the case of all symmetric $\zeta$ or even $\zeta$ such that $E[\zeta^3] \leq 0$). We leave this question open.

## 10 Proof of Proposition 1*b*

Because $Z = d\mathcal{P}^*/d\mathcal{P} = e^{-\omega(1)/2 - 1/8}$, the law of $Z$ is that of $\exp(Y^{-1/8, 1/4})$, where $Y^{-1/8, 1/4}$ is Normal variate with mean $-1/8$ and variance $1/4$. By standard calculations, then, the law of $Z$ has density function

$$f(y) = \sqrt{\frac{2}{\pi}} \frac{1}{y} \exp\left[-2\left(\ln(y) + \frac{1}{8}\right)^2\right].$$

Consider the function

$$V_0(y) = \frac{1}{f(y)} = \sqrt{\frac{\pi}{2}} \, y \exp\left[2\left(\ln(y) + \frac{1}{8}\right)^2\right]. \tag{10.1}$$

Differentiating $V_0$ shows that it is decreasing in $y$ as long as $\ln(y) < 3/8$, and a similar computation shows that $V_0$ is also convex on $(0, z_0)$ for small enough $z_0 > 0$. Hence, we may define a function $V : R_{++} \to R_{++}$ that coincides with $V_0$ on the interval $(0, z_0)$ and is extended to all of $(0, \infty)$ and that is convex, decreasing, differentiable, and satisfies $V'(\infty) = 0$. The conjugate function to this $V$, denoted $U$, therefore satisfies the conditions (1.1).

---

[7]Having a variance of 1 changes the formulas but not the basic conclusion.

We want to determine the range of values of strictly positive $y$ for which the value function
$$v(y) = \mathbb{E}_{\mathcal{P}}\big[V(y\,Z))\big] \tag{10.2}$$
is finite and for which values it is infinite. Clearly, this depends only on the small values of $Z = d\mathcal{P}^*/d\mathcal{P}$. Write the expectation (10.2) over the interval where $Z = d\mathcal{P}^*/d\mathcal{P} \in (0, z_0)$ as

$$\mathbb{E}_{\mathcal{P}}\big[V(yZ(\omega)); Z(\omega) \in (0, z_0)\big] = \int_0^{z_0} V_0(yz) f(z) dz$$
$$= \int_0^{z_0} \frac{1}{f(yz)} f(z) dz = \frac{1}{y} \int_0^{yz_0} \frac{f(w/y)}{f(w)} dw,$$

where the last step involves the change of variable $w = yz$.

By straightforward calculation, we find that, for some constant $K$ depending on $y$,
$$\frac{f(w/y)}{y f(w)} = K w^{4\ln(y)}; \text{ hence, } \mathbb{E}_{\mathcal{P}}\big[V(yz); Z \in (0, z_0)\big] = K \int_0^{yz_0} w^{4\ln(y)} dw.$$

This integral diverges if $4\ln(y) \leq -1$; that is, if $y \leq e^{-1/4}$.

By the duality between $u$ and $v$, this demonstrates, for $y_0 = e^{-1/4}$, the possibility that $v(y) = \infty$ for $y \leq y_0$ and is finite for $y > y_0$; moreover, as $y \searrow e^{-1/4} = y_0, v(y) \nearrow \infty$. This is possibility $b(i)$ in Proposition 1. And to extend this result to a general $y_0 > 0$, it suffices to pass from the function $V_0(y)$ to
$$V^{y_0}(y) = V\left(\frac{e^{-1/4}}{y_0} y\right).$$

For possibility $b(iii)$ in Proposition 1: Lemma 2 shows that
$$\mathbb{E}_{\mathcal{P}}\big[(y\,Z)^{-\alpha}\big] = e^{(\alpha^2+\alpha)/8} y^{-\alpha}.$$

Recall that $\phi(\alpha) := e^{(\alpha^2+\alpha)/8}$. Begin by defining, for $y > 0$,
$$V(y) := \sum_{k=1}^{\infty} \beta_k y^{-\alpha_k}, \tag{10.3}$$

where $\alpha_k$ and $\beta_k$ are given by
$$\alpha_k = 2^k - 1 \quad \text{and} \quad \beta_k = \frac{1}{2^k \alpha_k \phi(\alpha_k)}.$$

The sum (10.3) converges for all $y > 0$ and, in fact, does so faster than geometrically past some $k_0$ (that depends on $y$).

Moreover, it is evident that $V$ is strictly positive, convex, and twice (and more) continuously differentiable. And from Lemma 3, we have that
$$v(y) = \mathbb{E}_{\mathcal{P}}\big[V(y\,Z)\big] = \sum_{k=1}^{\infty} \beta_k \mathbb{E}_{\mathcal{P}}\big[(y\,Z)^{-\alpha_k}\big] = \sum_{k=1}^{\infty} \beta_k \phi(\alpha_k) y^{-\alpha_k} \tag{10.4}$$

$$\text{and} \quad v'(y) = -\sum_{k=1}^{\infty} \beta_k \phi(\alpha_k) \alpha_k y^{-\alpha_k - 1}. \tag{10.5}$$

Substituting in the formulas for $\alpha_k$ and $\beta_k$, (10.4) and (10.5) become

$$v(y) = \sum_{k=1}^{\infty} \frac{y^{-2^k+1}}{2^k(2^k-1)} \quad \text{and} \quad v'(y) = -\sum_{k=1}^{\infty} \frac{y^{-2^k}}{2^k}. \tag{10.6}$$

By inspection, $v(y) = \infty$ for $y < 1$ and is finite for $y \geq 1$. And $v'(y)$ is finite for $y \geq 1$. This, then, is the possibility $b(iii)$ in Proposition 1, for $y_0 = 1$.

We leave to the reader the construction of an example of possibility $b(ii)$ and examples where the pole is $y_0 \neq 1$.